\documentclass[aps, prb, showpacs, preprint, amsmath, letterpaper]{revtex4-1}

\usepackage{graphics}
\usepackage{amssymb}
\usepackage{bm}

\usepackage[hmargin=2cm,vmargin=2.5cm]{geometry}

\usepackage[pdftex]{color}
\usepackage{verbatim}
\pdfoutput=1

\usepackage{booktabs,microtype,afterpage} 

\usepackage{siunitx}

\usepackage[charter,greekuppercase=italicized]{mathdesign}

\usepackage[colorlinks,plainpages=false,linkcolor=blue,urlcolor=blue,citecolor=blue,pdfpagemode=UseNone,pdfstartview=FitBH]{hyperref}
\usepackage[charter,greekuppercase=italicized]{mathdesign}
\definecolor{red}{rgb}{0.85,.1,0}
\definecolor{green}{rgb}{0.0,0.6,0.0}
\definecolor{orange}{rgb}{1,0.5,0}
\usepackage{graphicx}    
\graphicspath{{./}{figure/}}
\DeclareGraphicsExtensions{.eps,.png,.pdf,.jpg}

\begin{document}

\title{Kondo driven suppression of charge density wave in van der Waals material UTe$_3$}	

\author{Justin Shotton$^{1,\dagger}$, Jiahui Zhu$^{1,\dagger}$, David Martinez$^{1}$, Diana Golovanova$^{2}$, Dipanjan Chaudhuri$^{3,4}$, Xuefei Guo$^{3,4}$, Peter Abbamonte$^{3,4}$, Feng Ye$^{5}$, Yiqing Hao$^{5}$, Huibo Cao$^{5}$, Suk Hyun Sung$^{6}$, Carly Grossman$^{6}$, Ismail El Baggari$^{6}$, Gal Tuvia$^{7}$, Mengke Liu$^{7}$, Ruizhe Kang$^{7}$, Matt Boswell$^{8}$, Weiwei Xie$^{8}$, Debapratim Pal$^{9}$, Anil Kumar$^{9}$, Yun Suk Eo$^{9}$, Binghai Yan$^{10,*}$, Kai Sun$^{11,*}$, Jonathan Denlinger$^{12,*}$, Sheng Ran$^{1,*}$}

\affiliation{$^1$ Department of Physics, Washington University in St. Louis, St. Louis, MO 63130, USA
\\$^2$  Department of Condensed Matter Physics, Weizmann Institute of Science, Rehovot 7610001, Israel
\\$^3$ Department of Physics, University of Illinois, Urbana, IL 61801, USA
\\$^4$ Materials Research Laboratory, University of Illinois, Urbana, IL 61801, USA
\\$^5$ Neutron Scattering Division, Oak Ridge National Laboratory, Oak Ridge, TN 37831, USA
\\$^6$ The Rowland Institute at Harvard, Harvard University, Cambridge, MA 02138, USA
\\$^7$ Department of Physics, Harvard University, Cambridge, MA 02138, USA
\\$^8$ Department of Chemistry, Michigan State University, East Lansing, MI 48824, USA
\\$^9$ Department of Physics and Astronomy, Texas Tech University, Lubbock, TX 79409, USA
\\$^{10}$ Department of Physics, The Pennsylvania State University, University Park, PA 16802, USA
\\$^{11}$ Department of Physics, University of Michigan, Ann Arbor, MI 48109, USA
\\$^{12}$ Advanced Light Source, Lawrence Berkeley National Laboratory, Berkeley, CA 94720, USA
\\$\dagger$ Equal contribution
\\$^*$ Corresponding authors
}

\date{\today}

\begin{abstract}

Competing electronic instabilities lie at the heart of emergent phenomena in quantum materials. In low-dimensional metals, Fermi-surface nesting can drive charge density wave (CDW) formation through a Peierls-like mechanism, while in strongly correlated systems, Kondo hybridization reconstructs the electronic structure by entangling localized moments with itinerant electrons. How these two fundamentally different instabilities interact—whether they coexist, compete, or mutually exclude each other—remains an open question. Here, we present suppression of charge density wave via the Kondo interaction in van der Waals material UTe$_3$. The angle-resolved photoemission spectroscopy (ARPES) data reveals Fermi surface nesting under similar conditions as seen in RETe$_3$ compounds. Despite that, no CDW is found in UTe$_3$ after an extensive search. We demonstrate that strong hybridization between U 5$f$ electrons and Te $p$ states reconstructs the low-energy electronic structure, removes the instability, and preempts CDW formation. Our results reveal a rare example where Kondo hybridization preempts density wave formation, offering a new route to controlling ordering phenomena in correlated 2D materials.

\end{abstract}

\maketitle
\section{Introduction}


The interplay between charge density waves (CDWs) and other electronic orders in two-dimensional (2D) materials unveils a rich spectrum of novel quantum phenomena critical to advances in quantum materials. In high-$T_c$ cuprate superconductors, for example, CDWs emerge at high temperatures and intertwine with magnetic orders to form stripe phases that compete with superconductivity\cite{lee2021xray}. In certain cases, their coexistence gives rise to more exotic phases such as pair density waves, where the superconducting order parameter itself develops a spatial modulation\cite{agterberg2020physics,liu2023pair,gu2023detection,chen2021roton,cao2024directly}.  CDWs have also been reported to coexist with antiferromagnetic orders, facilitating the emergence of chiral magnetic states and anomalous Hall effects\cite{teng2022discovery, yang2020giant, yu2021concurrence, wang2021field, ghimire2020competing}.

Another intriguing frontier lies in understanding how CDWs interact with Kondo lattice. In Kondo systems, conduction electrons hybridize with localized moments to form a many-body singlet state, fundamentally altering the low-energy electronic structure. Theoretical studies suggest that charge ordering can emerge from 2D Kondo lattices or 1D Kondo chains~\cite{huang2019charge, misawa2013charge, peters2013charge}. Indeed, recent experiments have observed the coexistence of CDWs and Kondo resonance in 2D materials~\cite{liu2024charge, luo2025direct}. Notably, these systems are based on $d$-electron Kondo physics, where the hybridization is relatively weak.

In principle, however, Kondo interactions and CDW formation can be competitive. CDWs are determined by the real part of the electronic susceptibility $\chi'(\mathbf{q})$~\cite{Mazin2008}, which reflects how efficiently electrons can lower their energy by developing a density modulation at momentum $\mathbf{q}$. The structure of $\chi'(\mathbf{q})$ depends sensitively on the band dispersion, the distribution of electronic states, and the available scattering channels across the entire Brillouin zone. The formation of a Kondo lattice reorganizes these ingredients~\cite{checkelsky2024flat}: hybridization reconstructs the Fermi surface, opens hybridization gaps, and introduces additional scattering processes between light conduction electrons and localized f-states. These effects broaden the electronic states, smear sharp features in $\chi'(\mathbf{q})$, and generally frustrate CDW instabilities. Moreover, Kondo hybridization promotes the formation of heavy, coherent quasiparticles at low temperatures, thereby suppressing the electronic and lattice fluctuations that often drive CDW transitions. Despite this expected competition, direct experimental evidence of Kondo-induced suppression of CDW has remained elusive.

The RETe$_3$ compounds are widely regarded as model systems for studying CDWs in layered materials. With only a few exceptions---radioactivity (RE = Pm), crystal instability (RE = Lu, Yb), or divalence (RE = Eu)---all members of the family host a CDW instability\cite{Yumigeta2021,Gweon1998,DiMasi1995,Ru2008,Banerjee2013}. Several (RE = Tb--Tm) even support a second CDW along an orthogonal crystallographic axis. The transition temperature of the primary CDW decreases monotonically with increasing rare-earth mass, ranging from above 600~K for RE = La to $\sim 250$~K for RE = Tm. In all cases, the primary CDW is a unidirectional, incommensurate order along the $c$-axis, one of the in-plane axes, characterized by a wave vector $q_c \approx (2/7)c^*$\cite{Yumigeta2021,Ru2008}. This vector matches a nested portion of the Fermi surface, supporting a Peierls-like mechanism for CDW formation, although the nesting condition is not perfect. In those compounds that exhibit a second CDW, the order sets in at a lower temperature with a wave vector $q_a \approx (1/3)a^*$, the other in-plane axis, perpendicular to $q_c$. This secondary CDW is likely associated with residual nesting features of the Fermi surface that remain after the primary CDW gap has opened\cite{Yumigeta2021,Banerjee2013}. 

In this context, we investigate UTe$_3$, a uranium-based analog of RETe$_3$, and present compelling evidence that Kondo hybridization suppresses CDW formation. Unlike RETe$_3$, no CDW has been detected in UTe$_3$ by transmission electron microscopy (TEM), X-ray diffraction, or neutron scattering. This absence is striking given that the dispersive, non-$f$ Fermi surface sheets of UTe$_3$ closely resemble the Te-derived Fermiology of the RETe$_3$, including clear nesting-like features. Crucially, ARPES measurements reveal a Kondo band near the Fermi level in UTe$_3$, absent in the rare-earth counterparts. Our tight binding model shows that introducing Kondo hybridization between the localized $f$ states and the dispersive bands reshapes the low-energy Fermi surface and electronic susceptibility. In combination with the additional scattering channels provided by this heavy quasiparticle band, the reconstructed electronic structure naturally suppresses the CDW instability. These findings reveal a rare instance of Kondo-driven suppression of CDW and broaden the landscape of competing orders in quantum materials.




\section{Absence of Charge Density Wave}

UTe$_3$ crystallizes in the same orthorhombic space group Cmcm (No. 63) as other RETe$_3$ compounds. It exhibits a larger in-plane lattice anisotropy, with lattice constants $a$ = $c$ = 4.35 Å, and a large out-of-plane lattice parameter $b$ = 24.79 Å, which spans two formula unit slabs of Te–U–Te along the stacking direction.

\begin{figure*}
    \centering
    \includegraphics[width=1\linewidth]{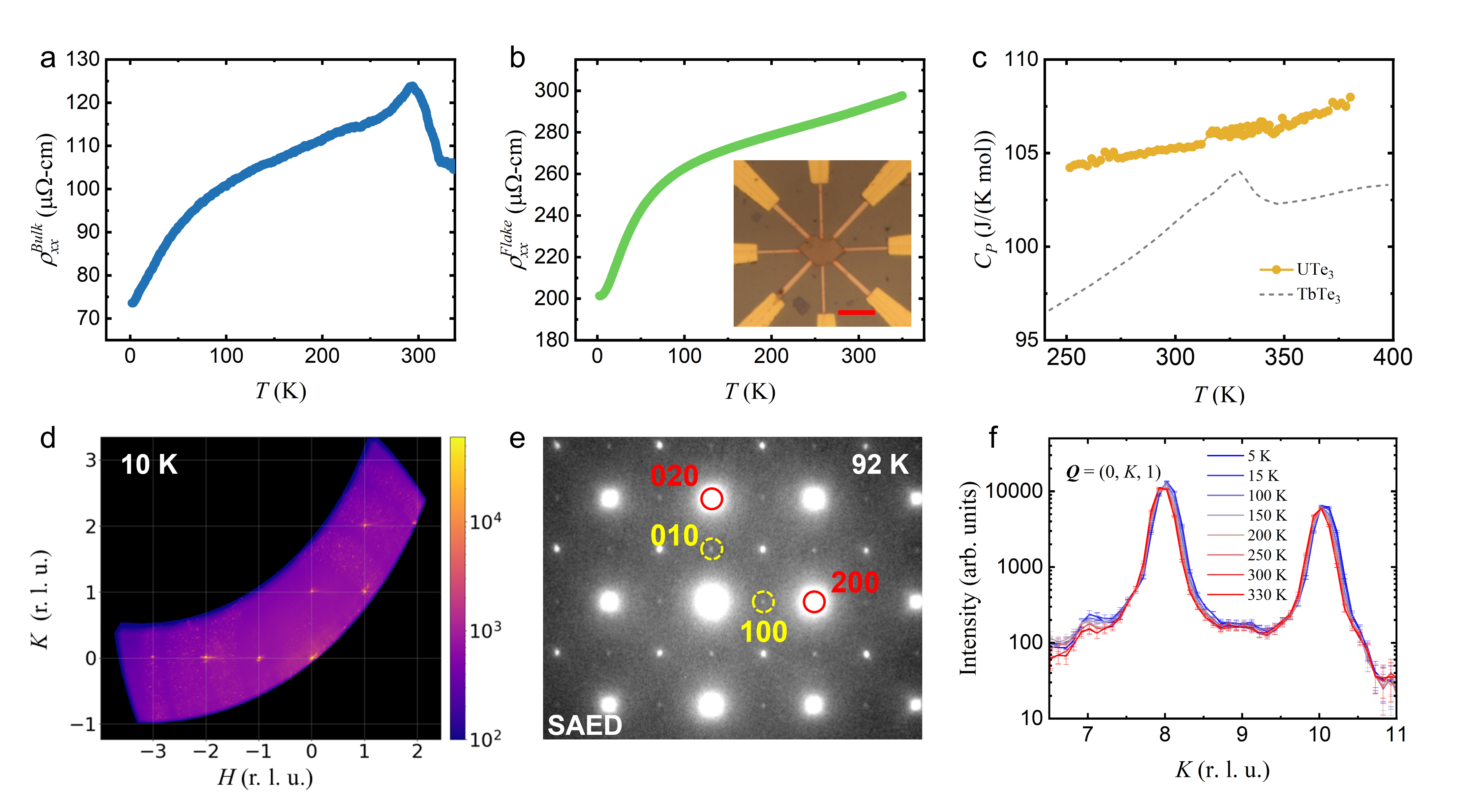}
    \caption{\textbf{Characterization of UTe$_3$} (a) Resistivity of a bulk sample of UTe$_3$. (b) Resistivity of a thin flake of UTe$_3$. The inset shows the flake stacked on top of Cr/Au bottom contacts. The red scale bar is 30 $\mu$m. (c) Heat capacity of UTe$_3$ from 250 to 380 K. The dashed line is the heat capacity of TbTe$_3$\cite{ma2025thermoelectric}, showing its CDW transition as comparison. (d) X-ray diffraction of a single crystal sample of UTe$_3$ at 10 K. (e) In-plane TEM diffraction pattern of UTe$_3$ at 92~K. (f) Neutron diffraction rocking curve near Q = (0, K, 1) measured various temperatures.}
    \label{fig:fig1}
\end{figure*}

Fig.\ref{fig:fig1}a shows the temperature-dependent resistivity of UTe$_3$ in a bulk sample. A clear signature of a phase transition is observed near 320 K, where the resistivity exhibits a sudden increase upon cooling. This behavior is similar to that of other RETe$_3$ compounds\cite{Ru2008}, in which the resistivity upturn is associated with the formation of a charge density wave that gaps out portions of the Fermi surface. However, unlike RETe$_3$~\cite{ma2025thermoelectric}, no corresponding anomaly is observed in the specific heat data of UTe$_3$, indicating that this is not a bulk thermodynamic phase transition. Furthermore, in a thin flake of UTe$_3$ (50 nm thick), the resistivity shows a smooth, monotonic decrease with no indication of a transition near 320 K.

These observations suggest that the resistivity anomaly in the bulk sample may not originate from intrinsic electronic ordering such as CDW. Instead, it could result from interlayer effects—such as sliding between van der Waals layers upon cooling—a phenomenon that has been reported in other layered materials\cite{philip2023local,cha2023order,zhang2024stacking}. Such effects are expected to be minimized or entirely suppressed in thin flakes, consistent with the absence of the anomaly in the 50-nm-thick sample.



Although these results argue against the presence of a CDW in UTe$_3$, it remains possible that a weak or short-range CDW could produce features too subtle to be resolved in transport or thermodynamic measurements. To more directly probe for such subtle signatures, we performed single-crystal X-ray diffraction (XRD), transmission electron microscopy (TEM), and neutron diffraction measurements. In other RETe$_3$ compounds, CDW satellite peaks were clearly visible in these measurements\cite{DiMasi1995,Banerjee2013}. Fig.\ref{fig:fig1}d shows the single-crystal XRD patterns of UTe$_3$ — specifically, $H$–$L$ momentum maps summed over the range $K = 19.9$ to $K = 21.1$—taken at 10~K. The diffraction pattern is indexed to an orthorhombic crystal structure 
with no satellite reflections characteristic of a CDW observed. Consistently, both in-plane (Fig.\ref{fig:fig1}e) and cross-sectional (Supplementary Fig. 12) TEM images show no evidence of satellite peaks along any crystallographic direction. Furthermore, single-crystal neutron diffraction measurements performed on 0.1 gram co-aligned single crystals while cooling from 330 K to 5 K along the $(0, K, 1)$ direction also reveal no satellite peaks (see the $0KL$ plot in Fig. 14 in Supplementary).

Together, our results provide strong evidence for the absence of a charge density wave in UTe$_3$. While we cannot entirely rule out the existence of an extremely weak CDW order that lies below our current detection limits, any such order would be significantly suppressed compared to the well-established CDW in RETe$_3$. Higher-resolution techniques may be required to fully explore this possibility, but even in that case, the dramatic contrast with RETe$_3$ underscores the strong suppression of CDW in UTe$_3$.


\section{Band Structure from the Te Square-Net Layer}


The absence of a CDW in UTe$_3$ is particularly surprising given that all known RETe$_3$ compounds exhibit robust CDW order. In RETe$_3$, the CDW is widely understood as a Peierls-type instability closely tied to the Fermi surface topology, as confirmed by both first-principles calculations and angle-resolved photoemission spectroscopy (ARPES)\cite{Brouet2008,seong2021angle,Gweon1998,Zong2018,Moore2010,Brouet2004,Regmi2023,Rettig2016,Rettig2013}. To investigate why UTe$_3$ deviates from this trend, we next turn to its electronic structure and Fermi surface.

\begin{figure*}[tbh]
\begin{center}
\includegraphics[width=17cm]{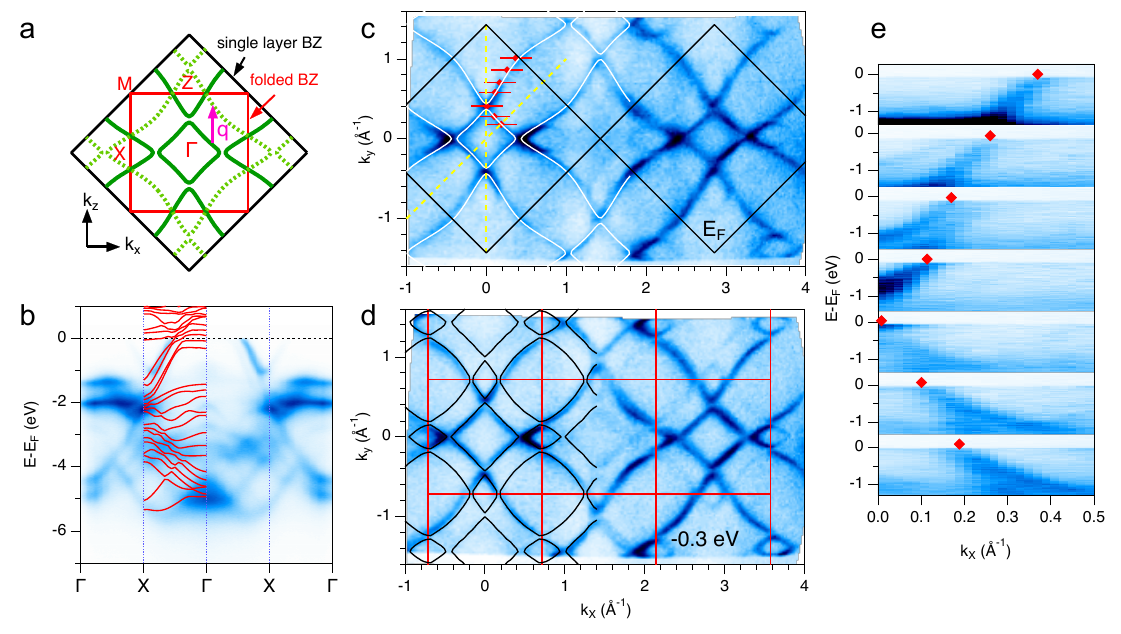} 
\caption{\textbf{Off-resonance ARPES and single Te-plane contribution to the electronic structure of UTe$_3$.} 
(a) Schematic of the single Te layer Fermi surface contours (solid) and BZ (black) compared to the bulk zone-folded BZ (red) with additional zone-folded FS contours (dashed).
(b) Wide valence band dispersion image along $X-\Gamma-X$ with comparison to DFT theory bands for bulk UTe$_3$.
(c) Wide FS map spanning multiple bulk Brillouin zones, measured at 92 eV at 10~K where U 5$f$ spectral weight is suppressed. 
Over-plotted are the BZ boundaries and constant energy contours from a single square-net Te plane. 
(d) Wide 92 eV map for a constant energy of -0.3 eV below $E_F$ that highlights zone-folded band intensities from the two Te-layer bulk structure.
(e) Series of energy band cuts at the (red) lines indicated in (c), illustrating the non-gapped $E_F$ crossings all along the Te layer FS contour.}
\label{fig3}
\end{center}
\end{figure*}

The full valence band dispersion along the $\Gamma - X$ direction, extending into the first and second Brillouin zones, is shown in Fig.\ \ref{fig3}b, measured using a 92~eV photon source. Both the hole-like and electron-like bands, derived primarily from Te 5$p$ orbitals, span a broad energy range of approximately 5 eV, with the hole-like bands crossing the Fermi level. Also overplotted is the full Te-$p$ band structure for an open-core UTe$_3$ (or ThTe$_3$) band calculation, which excludes U 5$f$ electrons. The excellent agreement between the ARPES intensity and the calculated band structure supports a Te-$5p$ character for these states. 



The corresponding Fermi surface map is shown in Fig.~\ref{fig3}c. In UTe$_3$ and other RETe$_3$, the two Te-layer have an A-B stacking, leading to zone folding, as illustrated schematically in Fig.~\ref{fig3}a. The Fermi surface can be viewed as two orthogonal, corrugated, quasi-1D hole-like pairs of Fermi surface contours derived from the Te square-net. These orthogonal bands hybridize and gap at their intersection points, forming a small diamond-shaped contour centered at the first reduced Brillouin zone and a larger curved diamond centered at the second reduced Brillouin zone. The dominant experimental Fermi surface contours agree well with the theoretical Fermi surface of a single Te square-net layer (overplotted), with the chemical potential adjusted to match the experimental band filling. The zone folding is more visible at a higher binding energy of -0.3 eV shown in Fig.~\ref{fig3}d. The comparison is made to theoretical contours of two Te square-net layers (overplotted), which highlights the formation of small eye-shaped (American football-shaped) contours that result from the intersection  of the zone-folded Fermi surface contours.




The quasi-1D character of the underlying hole-like Fermi surface shape, originating from the single Te square-net layer, naturally gives rise to a large electronic susceptibility and a nesting-like condition. A representative nesting vector is shown in Fig.\ \ref{fig3}a. In RETe$_3$, this nesting enhances the susceptibility at the CDW wave vector. Because the $k_F$ values vary along the Fermi surface, the nesting is imperfect, leading to a momentum-dependent partial gapping of the electronic structure~\cite{Gweon1998}. In contrast, no such gapping is observed in UTe$_3$, as shown in the Fermi surface map of Fig.\ \ref{fig3}c, consistent with the absence of CDW. This is further demonstrated in Fig.\ \ref{fig3}e, which displays a series of seven band dispersion images taken along horizontal momentum cuts from middle of the small diamond contour to the middle of the large diamond contour. In each case, the underlying band disperses to the Fermi level, and a clear Fermi edge is observed. Only in the regions near the crossing of the orthogonal quasi-1D contours, at the tips of the diamonds, is a hybridization gap seen below $E_F$, which is attributed to band interaction rather than CDW formation.


\section{U 5$f$ band structure and Kondo hybridization}



Although the band structure derived from the Te-$p$ states in UTe$_3$ closely resembles that of RETe$_3$, the presence of U 5$f$ electrons introduces new and significant elements to the electronic structure and underlying physics. To probe these U 5$f$ states, we tune the ARPES photon energy to 98 eV where the U 5$d$-to-5$f$ absorption threshold resonantly enhances the U 5$f$ spectral weight. We observed pronounced 5$f$spectral weight near the Fermi level, appearing in two distinctly different regions bounded by the Te double-layer Fermi surface contours and with different energies. 

\begin{figure*}[ht]
\begin{center}
\includegraphics[width=16cm]{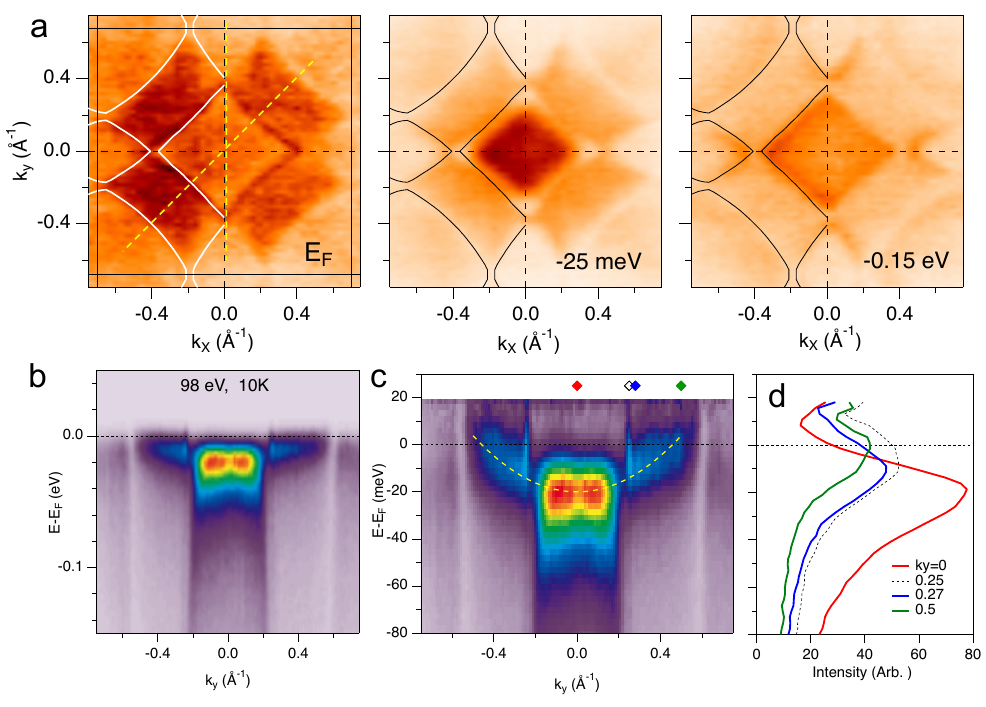} 
\caption{
\textbf{On-resonant ARPES electronic structure of U 5$f$ states and temperature dependence.} 
(a) Constant energy slices of a 98 eV map at 10K highlighting different confined regions of U 5f weight: outer rectangular regions at $E_F$, and a square zone-center region at -20 meV. The Te-$p$ dispersion band is highlighted at -0.15 eV.
Overplotted are constant energy contours from double-layer square-net Te planes with A-B stacking. 
(b) Near $E_F$ M-$\Gamma$-M band dispersion along the dashed diagonal cut in (a)
(c) Division of the data in (b) by the Fermi-Dirac function to enhance spectral weight above $E_F$, to highlight the shallow electron-like dispersion origin of the gapped states at the zone center. The dashed line parabola has en effective mass of m*=40.
(d) Line spectra at different momenta in (c) also illustrating the electron-like dispersion.  
}
\label{fig4}
\end{center}
\end{figure*}

In Fig.~\ref{fig4}a, the Fermi surface map at 98~eV exhibits high intensity in rectangular regions between the small diamond contour—originating from the single Te-layer band—and the zone-folded large diamond contour.  Notably, spectral weight is largely excluded from both the interior of the small diamond and the eye-like electron regions. A sharp small-diamond contour from Te-$p$ state remains clearly visible, similar to that seen in the off-resonance 92~eV map. However at -25 meV below $E_F$, strong 5$f$ spectral weight is confined to the interior of the small diamond FS contour. The corresponding band dispersion along a diagonal momentum cut (dashed line) is shown in Fig.~\ref{fig4}b. It highlights the relative strength of the gapped 5$f$ states at the BZ center and the Fermi-level-touching 5$f$ states in the outer rectangular regions. These two regions are separated by the fast-dispersing, hole-like Te-$p$ band. 


Closer inspection of the 5$f$ spectral weight suggests the presence of an underlying shallow electron-like dispersion that extends through both regions. To highlight this electron-like dispersion better, we divide the dispersion image in Fig.~\ref{fig4}b by a resolution-convolved Fermi-Dirac distribution, thereby enhancing the spectral weights above $E_F$. As shown in Fig.~\ref{fig4}c, the outer edges exhibit stronger intensity above $E_F$, enabling a shallow parabolic dispersion to be traced, with effective mass of 40. Complementary line spectra at three points along the heavy band are shown in Fig.~\ref{fig4}d, illustrating the electron-like peak shifting and relative intensities. Notably, the Fermi momentum $k_F$ of this heavy $f$ band nearly coincides with the $k_F$ of the zone-folded outer large Te-$p$ electron sheet. This apparent momentum-space overlap suggests Kondo hybridization between the itinerant Te-$p$ and the heavy $f$ bands, which is central to understanding the suppression of the CDW formation in UTe$_3$, as will be further discussed below.



\begin{figure*}[ht]
\begin{center}
\includegraphics[width=16cm]{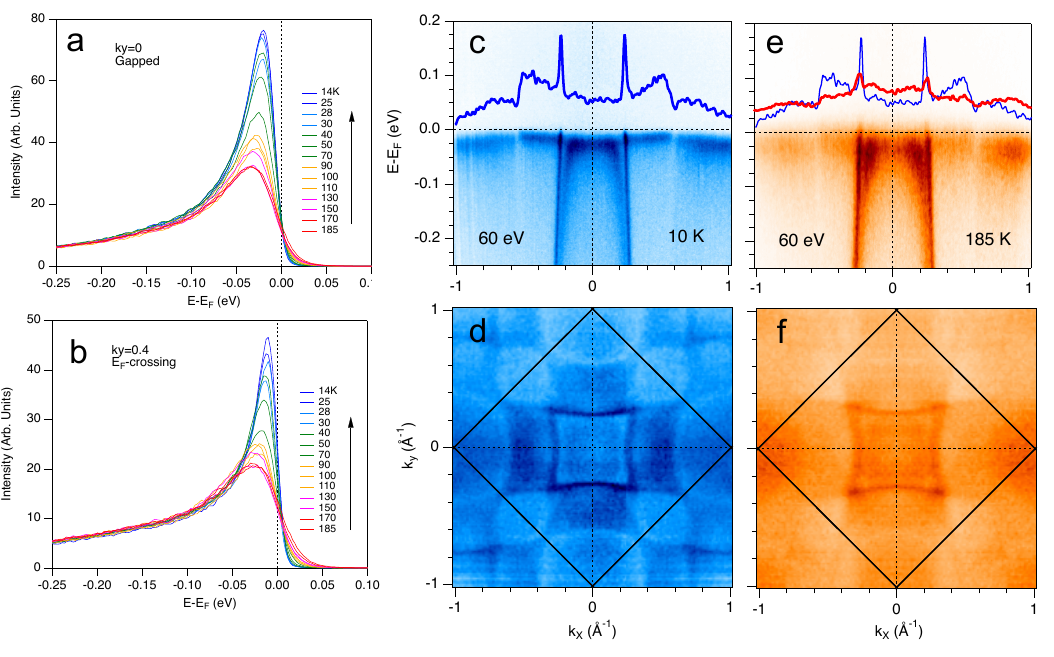} 
\caption{
\textbf{On-resonant ARPES electronic structure of U 5$f$ states.} 
(a,b) Temperature-dependent on-resonance 98 eV U 5$f$ spectral weight in the two different regions showing similar decoherence behavior. The strong 5$f$ weight at high temperature gives evidence for a high Kondo temperature. 
Comparison of 10~K and 185~K band dispersions (c,e) and Fermi surface maps (d,f) measured at 60 eV using LH polarization, where the U 5$f$ heavy electron band and light mass Te 5$p$ intensities are comparable allowing their interaction to be probed.  The crystal is rotated 45 deg relative to the maps in Fig. 3 and Fig. 4. The T-dependence decoherence of the shallow electron band results in loss of spectral weight relative to the light Te 5$p$ band is observable in the band dispersion image and $E_F$ momentum intensity profile and in the FS map images. 
}
\label{fig5}
\end{center}
\end{figure*}

To further reveal the Kondo character of the U 5$f$ band near the Fermi level, we investigated temperature-dependence. As shown in Fig.~\ref{fig5}a and b, these on-resonance 98~eV measurements were performed by cleaving the sample at high temperature and cooling down to low temperature. A strong increase is observed in the U 5$f$ spectral weight upon cooling, both in the gapped central region and in the outer regions where the heavy electron band crosses the Fermi level, consistent with the formation of a coherent Kondo lattice. The U 5$f$ weight at 185~K still remains relatively large, about 40$\%$ of the 10~K intensity, indicating a large Kondo temperature, e.g. strong hybridization mixing of U $f$-states with a dispersive bands. 


The decrease in U 5$f$ spectral weight at high temperature is also visible in the band dispersion images and Fermi surface maps shown in Fig.~\ref{fig5}(c-f), measured at 60~eV photon energy with linear horizontal (LH) polarization. At this photon energy, the relative intensity of the Te 5$p$ and U 5$f$ bands is intermediate between the extremes of the off-resonance (92~eV) data in Fig.~2 and the on-resonance (98~eV) data in Figs.~3 and 4(a,b). This allows a qualitative visual observation of the decrease in the heavy electron band intensity between 10 K and 185 K in Figs.~\ref{fig5}c and e, and also the suppression of the outer Fermi surface regions of the heavy electron band in Figs.~\ref{fig5}d and f. A more quantitative comparison is presented in the momentum distribution curves (MDCs) at $E_F$ in Figs.~\ref{fig5}c and e. At 10~K, the MDC shows a sharp and narrow peak corresponding to the Te 5$p$ band crossing the Fermi level, along with an asymmetric rise in intensity toward the momentum region where the heavy electron band crosses $E_F$. At 185~K, the outer U 5$f$ spectral weight is strongly reduced, while the Te 5$p$ peaks still exists with a twofold width broadening and a more than threefold amplitude reduction. Note that the interior U 5$f$ spectral weight actually appears increased at 185 K due to the high temperature broadening of the U 5$f$ peak.

Also at 60~eV, the full Te 5$p$ band dispersion at higher binding energies becomes clearly visible, allowing the detection of a distinct velocity discontinuity in the band dispersion across the heavy electron band crossing. As quantified in the supplemental figure (Supplementary Fig. 5), the very light Te 5$p$ band exhibits a velocity of $\sim$7.5 eV-\AA\ below the heavy electron band. This velocity is renormalized by a factor of $\sim$4 to a much heavier slope of 1.7 eV-\AA\ above the heavy band at 10~K. This renormalization arises from the hybridization between the U 5$f$ states and the Te 5$p$ band near $E_F$, consistent with the band dispersion observed at 98~eV.


\section{Kondo driven suppression of the charge density wave} 

To investigate the impact of Kondo hybridization on the stability of the charge density wave, we employ a low-energy tight-binding model that captures the essential electronic degrees of freedom near the Fermi level. In this model, the dispersive bands originating from non-local orbitals—including both Te 5$p$ states and uranium-derived conduction states—are taken from first-principles calculations and reproduced using a tight-binding parametrization.

In addition to these dispersive bands, we include a weakly dispersive heavy electron band near the Fermi energy, motivated directly by our ARPES observations. Both ARPES measurements and DFT calculations indicate that this shallow electron band possesses mixed U $d$ and U $f$ character, and exhibits a finite bandwidth on the order of tens of meV with well-defined Fermi momenta distinct from those of the Te $p$ bands. We therefore model this band with a parabolic dispersion, whose parameters are extracted from the experimental data in Fig.~\ref{fig4}c.

Although this heavy band originates from hybridized U $d$–U $f$ states rather than a strictly localized $f$ orbital, its inclusion effectively captures the low-energy consequences of Kondo lattice physics relevant for the reconstruction of the Fermi surface. Higher-energy $f$ states and additional orbital complexity are neglected for simplicity, as they do not qualitatively affect the low-energy Fermi surface topology discussed here.

Within this framework, the Hamiltonian can be written as the following $5\times5$ matrix:
\begin{equation}
\hat{H}=\left(\begin{array}{ccc}
\hat{H}_h & \hat{H}_{\mathrm{hA}} & \hat{H}_{\mathrm{hB}}\\
 \hat{H}_{\mathrm{hA}}^\dagger & \hat{H}_{\mathrm{A}} & \hat{H}_{\mathrm{AB}} \\
\hat{H}_{\mathrm{hB}} ^\dagger&\hat{H}_{A B}^{\dagger} & \hat{H}_B
\end{array}\right)
\end{equation}
Here, the lower-right $4\times4$ block—consisting of $\hat{H}_{\mathrm{A}}$, $\hat{H}_{\mathrm{B}}$, and $\hat{H}_{\mathrm{AB}}$—describes the Te $p_x$ and $p_z$ orbitals that dominate the Fermi surface in RETe$_3$. $\hat{H}_h$ represents the weakly dispersive heavy electron band with mixed U $d$–U $f$ character. 

The hybridization between shallow heavy-electron band and Te $p$ orbitals must respect the crystal symmetries. Although this band is not a purely localized $f$ orbital, its hybridization with Te $p$ states is dominated by the $f$-orbital symmetry components. Therefore, we adopt a simplified symmetry-based treatment by assigning the dominant symmetry of the shallow band to that of the $f_{z^3}$ orbital (or equivalently, Wannier states transforming under the same representation).
It is important to note that the qualitative conclusions are insensitive to this particular choice of orbital symmetry. Because of mirror symmetry, tunneling between the $f$-like band and the $p_z$ orbitals of the A sublattice (or the $p_x$ orbitals of the B sublattice) is forbidden, as these states possess opposite mirror parity. Therefore, the allowed symmetry channels yield
\begin{equation}
\hat{H}_{\mathrm{hA}} = \left( 2 i\, t_{hp} \sin (\vec{k}\cdot \vec{a}_1), 0\right), \qquad 
\hat{H}_{\mathrm{hB}} =\left(0, 2 i\, t_{hp} \sin (\vec{k}\cdot \vec{a}_3)\right),
\end{equation}
where $t_{hp}$ denotes the tunneling amplitude between the shallow heavy-electron band and the neighboring tellurium $p$ orbitals. We take $t_{hp}=20~\text{meV}$, much smaller than the Te–Te hopping amplitudes, reflecting the localized nature of the $f$ electrons.

\begin{figure*}[ht]
\begin{center}
\includegraphics[width=16cm]{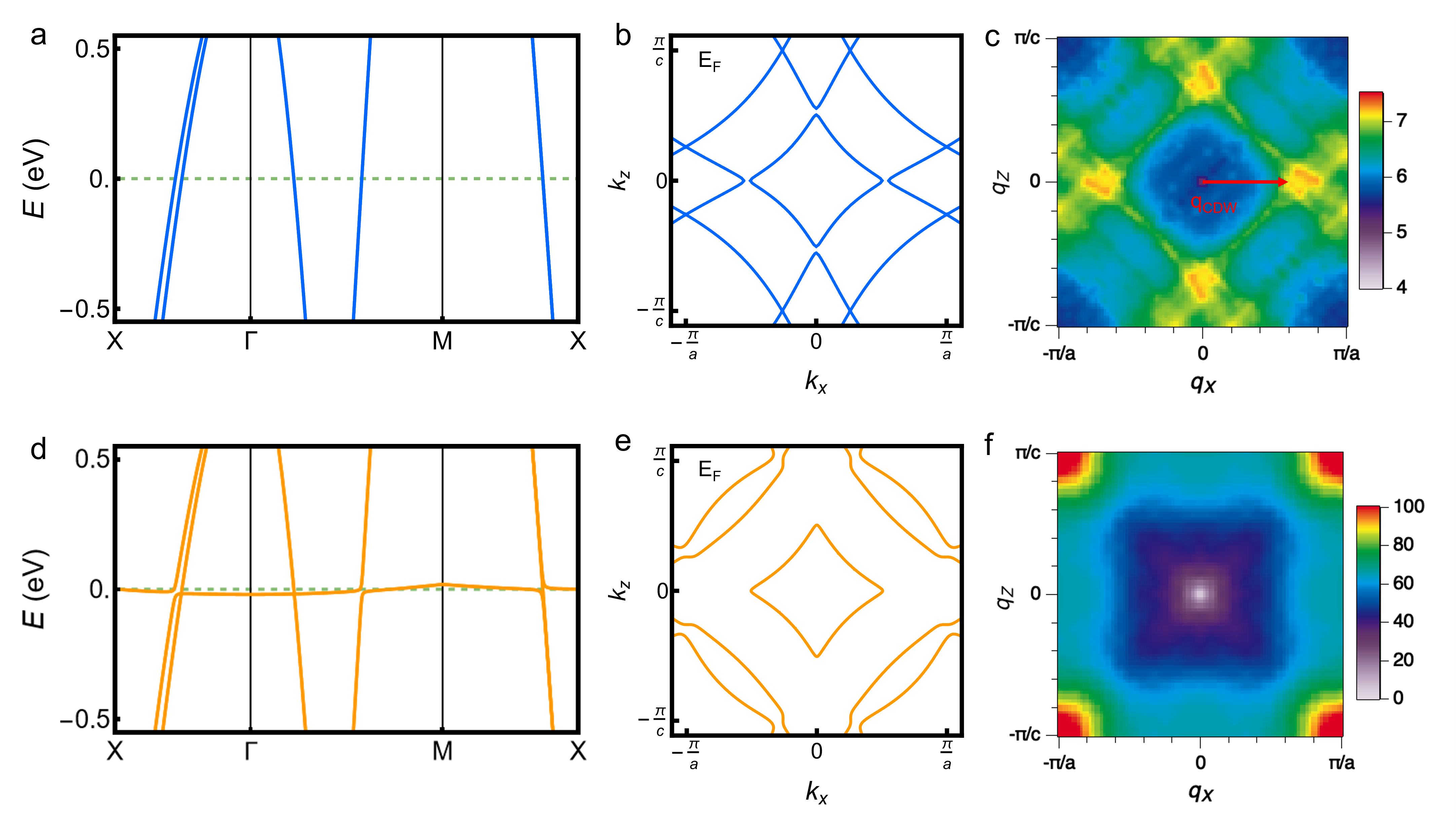} 
\caption{
\textbf{Tight binding model.} 
(a) Dispersion of the electronic bands near the Fermi energy without the $f$ band, where the horizontal green dashed lines indicate the Fermi energy. (b) The Fermi surface contours for the $p$ bands TB model. Without the $f$ states, the Fermi surface exhibits clear nesting, which favors CDW ordering. (c) The real part of the electronic susceptibility $\chi(q)$ for the $p$ bands TB model, exhibiting a clear peak responsible for CDW. (d, e, f) Similar to (a, b, c) only for the TB model that includes a shallow U $f$ band near the Fermi energy. Here, hybridization between the $f$ and $p$ bands introduces a significant reconstruction of the band structure near the Fermi energy, while the dispersions away from the Fermi level remain largely unchanged. Fermi surface nesting is modified, and the peak in electronic susceptibility is removed.
}

\label{TB}
\end{center}
\end{figure*}

Figure~\ref{TB} compares the band structure before and after inclusion of the heavy-electron band. Since the shallow $f$ band is pinned to the Fermi level by the Kondo resonance with only weak dispersion, its effects are confined to an energy window of order $\pm t_{hp}$ around the Fermi energy. As shown in Fig.~\ref{TB}d, hybridization between the $f$ and $p$ states strongly modifies the band structure near $E_F$, while dispersions away from the Fermi energy remain essentially unchanged. 

Similar to many other heavy-fermion compounds, this hybridization leads to a reconstruction of the Fermi surface: it opens a hybridization gap in one of the conduction bands and substantially reshapes the Fermi surface. This Kondo-coupling–induced Fermi-surface reconstruction, and its comparison to the $p$ bands without $f$ states, is illustrated in Fig.~\ref{TB}b and Fig.~\ref{TB}e, which show the Fermi surfaces without and with the inclusion of the $f$ band. The reconstructed Fermi surface in the presence of hybridization closely resembles the ARPES data shown in Fig.~\ref{fig4}a. A key consequence of the Kondo hybridization is the suppression of Fermi-surface nesting. This effect becomes particularly evident at energies slightly below the Fermi level, for example at -10~meV (Supplementary Fig. 8). 

Although the suppression of the nesting condition already suggests a weakening of the CDW instability, the formation of a charge density wave is ultimately governed by the real part of the electronic susceptibility rather than the nesting (related to the imaginary part of the electronic susceptibility). To conclusively assess the CDW tendency, we calculated the real part of the electronic susceptibility of UTe$_3$ within the tight-binding model, both without and with inclusion of the U $f$-derived band. The results are shown in Fig.~\ref{TB}c and f, respectively. 

For the tight-binding model containing only the Te $p$ bands, the susceptibility exhibits pronounced peaks at wave vectors near $q_{\mathrm{CDW}} \approx \pm 0.35,\pi/a$. This gives an imperfect incommensurate Fermi surface nesting condition, which favors CDW ordering. In contrast, upon inclusion of the $f$-derived band, the susceptibility is dramatically modified. The overall magnitude of the susceptibility in the model including the $f$ band is approximately an order of magnitude larger than that of the Te $p$-only model, indicating that the $f$-derived states dominate the low-energy response. The momentum dependence becomes significantly more uniform, with the previously pronounced nesting ridge largely suppressed. The intensity instead redistributes toward the Brillouin-zone corners, and no clear directional enhancement remains at the original CDW wave vector. This qualitative change indicates that Kondo hybridization effectively removes the instability responsible for CDW formation.

\section{Ferromagnetic ground state}

Not only is the charge density wave suppressed in UTe$_3$, but its magnetic ground state also differs markedly from that of all other RETe$_3$ compounds. UTe$_3$ exhibits a ferromagnetic phase that emerges near 20~K. Intuitively, this is consistent with the absence of a CDW-induced partial Fermi surface gap. The ungapped Fermi surface in UTe$_3$ retains a high density of states at the Fermi level, which favors ferromagnetism according to the Stoner model.

\begin{figure*}
    \centering
    \includegraphics[width=1\linewidth]{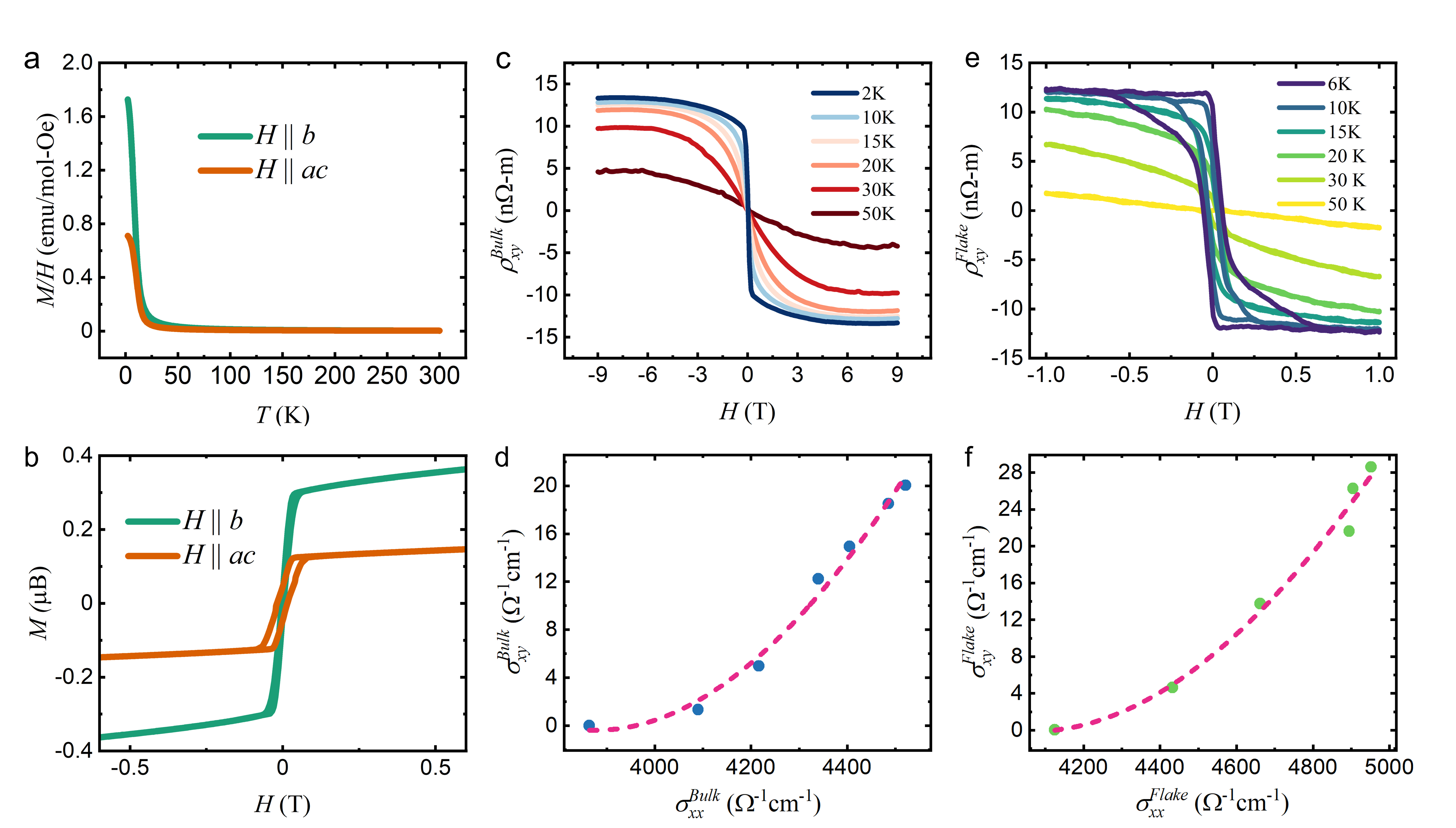}
    \caption{\textbf{Ferromagnetic Phase in UTe3} (a) Magnetic susceptibility of UTe$_3$. (b) Magnetization of UTe3. (c) Anomalous Hall effect in a bulk sample of UTe3. (d) \(\sigma^{Bulk}_{xy}\) as a function of \(\sigma^{Bulk}_{xx}\) with the fit \(\sigma^a_{xy}=a\sigma^2_{xx}+b/\sigma_{xx}+c\) to extract the intrinsic contribution. The intrinsic contribution was found to be approximately 850 /$\Omega$cm. (e) The same as (c) but with a thin flake sample. (f) The same as (d) but with a thin flake sample. The intrinsic contribution was found to be approximately 670 /$\Omega$cm. }
    \label{fig:ferro}
\end{figure*}

As shown in Fig. \ref{fig:ferro}a, the magnetic susceptibility reveals substantial components both in-plane and out-of-plane, indicating that the ferromagnetic moment is canted. The field-dependent magnetization displayed in Fig. \ref{fig:ferro}b exhibits clear hysteresis loops for both field orientations. The ordered moment is approximately 0.3~$\mu_B$ along the out-of-plane direction and 0.1~$\mu_B$ in-plane.

Figures~\ref{fig:ferro}c and d present the Hall resistivity of both bulk and thin-flake samples, each displaying a clear anomalous Hall effect within the ferromagnetic phase. The thin-flake sample exhibits a pronounced hysteresis loop, which persists up to 30~K, whereas the hysteresis loop in the bulk sample is much weaker and vanishes above 10~K. This indicates an enhancement of ferromagnetism in the thin flake regime, consistent with another recent work\cite{Thomas2025}. To disentangle the intrinsic Berry curvature contribution from extrinsic scattering mechanisms in the anomalous Hall effect, we performed a scaling analysis. As shown in Fig.~\ref{fig:ferro}e, the Hall conductivity is plotted against the longitudinal conductivity and fitted using the empirical relation \(\sigma^a_{xy}=a\sigma^2_{xx}+b/\sigma_{xx}+c\), where the constant term $c$ represents the intrinsic contribution. From this fit, we extract an intrinsic Hall conductivity of approximately 670 /$\Omega$cm for the flake sample, indicating a substantial Berry curvature effect in the anomalous Hall response.


\section{Conclusions}

The interplay between charge density waves and other electronic orders has been a central theme in the study of quantum materials for decades. While extensive work has explored the coexistence and competition between CDWs, superconductivity, and magnetism, the interaction between CDWs and Kondo hybridization remains much less understood. Coexistence of CDWs and Kondo hybridization has been observed in systems where Kondo coupling is relatively weak—either in materials dominated by d-electron Kondo physics or in cases where the Kondo resonance is largely suppressed at the Fermi level. In this work, we demonstrate that strong Kondo hybridization, as realized in the 5$f$-electron system UTe$_3$, can actively suppress CDW formation. Our ARPES measurements reveal the emergence of a Kondo band near the Fermi level, and our theoretical modeling shows that the hybridization between U 5$f$ and dispersive bands strongly reconstructs the Fermi surface and electronic susceptibility. This finding uncovers a new form of competition in correlated materials. Such suppression may not only eliminate charge order but also enable other exotic phases to emerge, including unconventional magnetism or superconductivity. Our results suggest that tuning the strength of Kondo hybridization—via pressure, chemical substitution, dimensionality, or interface engineering—could serve as a powerful tool for controlling electronic orders in low-dimensional quantum materials.

\section*{Methods}

\subsection{Crystal synthesis and characterization} 
Single crystals of UTe$_3$ were synthesized by the Molten Metal Flux method. The elements were mixed in the ratio U:Te = 1:15. The crucible containing the starting elements was sealed in a fused silica ampule and then heated up to 800ºC in a box furnace. The crystals grew as the temperature was reduced to 520ºC over 120 hours, after which the ampule was quickly removed from the furnace and the flux was decanted in a centrifuge. The crystal structure was determined by $x$-ray powder diffraction using a Rigaku $x$-ray diffractometer with Cu-K$_{\alpha}$ radiation. Several single crystals were picked up and examined in the Bruker D8 Quest Eco single-crystal X-ray diffractometer equipped with Photon II detector and Mo radiation ($\lambda_{K\alpha}$ = 0.71073 Å) to obtain the structural information. The crystal was measured with an exposure time of 10 s and a scanning 2$\theta$ width of 0.5° at room temperature. The data was processed in Bruker Apex III software and the structural refinement were conducted with the SHELXTL package using direct methods and refined by full-matrix least-squares on $F^2$. Electrical transport measurements were conducted using a standard lock-in technique, magnetization was measured with a vibrating sample magnetometer (VSM), and specific heat measurements were performed, all within a Quantum Design Physical Property Measurement System (PPMS)

\subsection{Single crystal X-ray diffraction} 

X-ray diffraction measurements of UTe$_3$ single crystals were carried out using a lab-based, four-circle Huber diffractometer equipped with a Mo K$\alpha$ ($\lambda$ = 0.7093 ) micro-focus x-ray source (Xenocs GeniX3D) and a Mar345 image plate detector. A closed-cycle helium cryostat with Be domes enclosure was mounted on the diffractometer for low-temperature diffraction experiments.

\subsection{ARPES} 
Angle-resolved photoelectron spectroscopy (ARPES) measurements were performed at the MERLIN beamline 4.0.3 of the Advanced Light Source (ALS) in the energy range of 30-150 eV employing both linear horizontal (LH) and linear vertical (LV) polarizations from an elliptically polarized undulator. 
A Scienta R8000 electron spectrometer was used in combination with a six-axis helium cryostat goniometer in the temperature range of 10-200K with a total energy resolution of $\geq$15 meV and base pressure of $<$5$\times$10$^{-11}$ Torr. 
Single crystal platelets of UTe$_3$ were razorblade cut to $\sim$2x2 mm lateral size, top-post cleaved in UHV, and the flattest uniform regions probed with an $\sim$50 $\mu$m beam spot size.



\subsection{TEM}
Out of plane TEM measurements were performed with a JEOL 2100F Scanning Transmission Electron Microscope at an operation voltage of 200 kV. The sample was prepared with a ThermoFisher Scios 2 Dual Beam Focus Ion Beam (FIB). A lamella was extracted from a small chunk of the material, then attached to a FIB half TEM grid, and polished to electron transparency.

Annular dark-field scanning transmission electron microscopy (ADF-STEM) and selected area electron diffraction (SAED) were performed on Thermo Fisher Scientific (TFS) Themis Z G3 operated at 200 keV with Mel-build double tilt cryogenic air-free transfer holder. Cross-sectional TEM samples were prepared on Helios 660. Plan-view TEM samples were prepared by exfoliating UTe$_3$ flakes onto polydimethylsiloxane (PDMS) gel stamps and mechanically transferring on to SiN$_x$ TEM grids from Norcada using a home-built transfer system inside Argon-filled glove box.

\subsection{Neutron diffraction} 
Neutron diffraction measurements were performed at HB3A DEMAND at High Flux Isotope Reactor at Oak Ridge National Laboratory~\cite{chakoumakos2011four}. The experiment used Si (220) monochromator with the wavelength of 1.542 Å\cite{chakoumakos2011four}. The experiment used coaligned single crystals on a Si substrate. The coaligned sample has total mass of 0.1g and mosaicity of ±5°. The sample was mounted on the four-circle goniometer and cooled down to 5 K using a helium closed cycle refrigerator. The data reduction used ReTIA\cite{hao2023machine}. The symmetry analysis used Bilbao Crystallography Server\cite{aroyo2006bilbao}. The structure refinement used Fullprof\cite{rodriguez1993recent}.

\section*{acknowledgement}

We acknowledge fruitful discussions with J. Hoffman. Research at Washington University was supported by the National Science Foundation (NSF) Division of Materials Research Award DMR-2236528. J.S. acknowledges the NRT LinQ, supported by the NSF under grant no. 2152221. K.S. acknowledges National Science Foundation through the Materials Research Science and Engineering Center at the University of Michi-gan Award No. DMR-2309029. B.Y. acknowledges the financial support by the Penn State Materials Research Science and Engineering Center for Nanoscale Science (MRSEC) under National Science Foundation award DMR-2011839. D.C., X.G. and P.A. acknowledge funding support from the Center for Quantum Sensing and Quantum Materials, an Energy Frontier Research Center funded by the U.S. Department of Energy (DOE), Office of Science, Basic Energy Sciences (BES), under award no. DE-SC0021238, and from the EPiQS program of the Gordon and Betty Moore Foundation under Grant No. GBMF9452. This research used resources at the High Flux Isotope Reactor, a DOE Office of Science User Facility operated by the Oak Ridge National Laboratory. The beam time was allocated to DEMAND at HB-3A on proposal number IPTS-31703.1. Y.H. and H.C. acknowledge the support from U.S. DOE BES Early Career Award KC0402010 under Contract No. DE-AC05-00OR22725. G.T. was supported by the Israeli Council of Higher Education Quantum Technology Fellowship. S.H.S, C.G, and I.E acknowledges support from the Rowland Institutes at Harvard University. This work was carried in part through the use of Harvard Center for nanoscale Systems (CNS), and MIT.nano’s facilities.

\section*{Competing Interests}
The authors declare no competing interests.


\bibliographystyle{apsrev4-2}
\bibliography{UTe3.bib} 

\clearpage

\sectionmark{FIG. 4. U5f electron pocket}

\sectionmark{FIG. 5. T-Dependence}

\end{document}